\documentstyle{mn}

\newif\ifAMStwofonts
\AMStwofontstrue

\ifoldfss
  \ifCUPmtlplainloaded \else
    \NewTextAlphabet{textbfit} {cmbxti10} {}
    \NewTextAlphabet{textbfss} {cmssbx10} {}
    \NewMathAlphabet{mathbfit} {cmbxti10} {} 
    \NewMathAlphabet{mathbfss} {cmssbx10} {} 
  \fi
  \ifAMStwofonts
    \ifCUPmtlplainloaded \else
      \NewSymbolFont{upmath} {eurm10}
      \NewSymbolFont{AMSa} {msam10}
      \NewMathSymbol{\upi}     {0}{upmath}{19}
      \NewMathSymbol{\umu}     {0}{upmath}{16}
      \NewMathSymbol{\upartial}{0}{upmath}{40}
      \NewMathSymbol{\leqslant}{3}{AMSa}{36}
      \NewMathSymbol{\geqslant}{3}{AMSa}{3E}

       \let\le=\leqslant
       \let\ge=\geqslant
    \fi
  \fi
\fi 

\ifnfssone
  \newmathalphabet{\mathit}
  \addtoversion{normal}{\mathit}{cmr}{m}{it}
  \addtoversion{bold}{\mathit}{cmr}{bx}{it}
  \newmathalphabet{\mathbfit} 
  \addtoversion{normal}{\mathbfit}{cmr}{bx}{it}
  \addtoversion{bold}{\mathbfit}{cmr}{bx}{it}
  \newmathalphabet{\mathbfss} 
  \addtoversion{normal}{\mathbfss}{cmss}{bx}{n}
  \addtoversion{bold}{\mathbfss}{cmss}{bx}{n}
  \ifAMStwofonts
    \ifCUPmtlplainloaded \else
      \UseAMStwoboldmath
      \makeatletter
      \new@mathgroup\upmath@group
      \define@mathgroup\mv@normal\upmath@group{eur}{m}{n}
      \define@mathgroup\mv@bold\upmath@group{eur}{b}{n}
      \edef\UPM{\hexnumber\upmath@group}
      \new@mathgroup\amsa@group
      \define@mathgroup\mv@normal\amsa@group{msa}{m}{n}
      \define@mathgroup\mv@bold\amsa@group{msa}{m}{n}
      \edef\AMSa{\hexnumber\amsa@group}
      \makeatother
      \mathchardef\upi="0\UPM19
      \mathchardef\umu="0\UPM16
      \mathchardef\upartial="0\UPM40
      \mathchardef\leqslant="3\AMSa36
      \mathchardef\geqslant="3\AMSa3E

       \let\le=\leqslant
       \let\ge=\geqslant
    \fi
  \fi
\fi 

\ifnfsstwo
  \DeclareMathAlphabet{\mathbfit}{OT1}{cmr}{bx}{it}
  \SetMathAlphabet\mathbfit{bold}{OT1}{cmr}{bx}{it}
  \DeclareMathAlphabet{\mathbfss}{OT1}{cmss}{bx}{n}
  \SetMathAlphabet\mathbfss{bold}{OT1}{cmss}{bx}{n}
  \ifAMStwofonts
    \ifCUPmtlplainloaded \else
      \DeclareSymbolFont{UPM}{U}{eur}{m}{n}
      \SetSymbolFont{UPM}{bold}{U}{eur}{b}{n}
      \DeclareSymbolFont{AMSa}{U}{msa}{m}{n}
      \DeclareMathSymbol{\upi}{0}{UPM}{"19}
      \DeclareMathSymbol{\umu}{0}{UPM}{"16}
      \DeclareMathSymbol{\upartial}{0}{UPM}{"40}
      \DeclareMathSymbol{\leqslant}{3}{AMSa}{"36}
      \DeclareMathSymbol{\geqslant}{3}{AMSa}{"3E}

       \let\le=\leqslant
       \let\ge=\geqslant
    \fi
  \fi
\fi 

\ifCUPmtlplainloaded \else
  \ifAMStwofonts \else 
    \def\upi{\pi}
    \def\umu{\mu}
    \def\upartial{\partial}
  \fi
\fi

\title{A New Quadruple Gravitational Lens System: CLASS B0128+437}
\author[P. M. Phillips et al.]
       {P.~M.~Phillips$^1$, 
M.~A.~Norbury$^1$, 
L.~V.~E.~Koopmans$^{1,2}$,
I.~W.~A.~Browne$^1$,
\newauthor
N.~J.~Jackson$^1$,
P.~N.~Wilkinson$^1$,
A.~D.~Biggs$^1$,
R.~D.~Blandford$^3$,
\newauthor
A.~G.~de~Bruyn$^{4,2}$,
C.~D.~Fassnacht$^5$,
P.~Helbig$^{2,1}$,
S.~Mao$^1$,
D.~R.~Marlow$^6$,
\newauthor
S.~T.~Myers$^5$,
T.~J.~Pearson$^3$,
A.~C.~S.~Readhead$^3$,
D.~Rusin$^6$,
E.~Xanthopoulos$^1$\\
$^1$University of Manchester, Jodrell Bank Observatory, Macclesfield, Cheshire, SK11 9DL U.K.\\
$^2$Kapteyn Astronomical Institute, Postbus 800, NL-9700 AV Groningen, The Netherlands\\
$^3$California Institute of Technology, Pasadena, CA 91125, U.S.A.\\
$^4$Netherlands Foundation for Research in Astronomy, Postbus 2, NL-7990 AA Dwingeloo, The Netherlands\\
$^5$National Radio Astronomy Observatory, P.O. Box 0, Socorro, NM 87801, U.S.A.\\
$^6$Department of Physics and Astronomy, University of Pennsylvannia, 209 South 33rd Street, Philadelphia, PA 19104, U.S.A.}
\date{}

\pagerange{\pageref{firstpage}--\pageref{lastpage}}
\pubyear{2000}

\begin{document}

\maketitle

\label{firstpage}

\begin{abstract}
High resolution MERLIN observations of a newly-discovered four-image gravitational lens system, B0128+437, are presented. The system was found after a careful re-analysis of the entire CLASS dataset. The MERLIN observations resolve four components in a characteristic quadruple-image configuration; the maximum image separation is 542~mas and the total flux density is 48~mJy at 5~GHz. A best-fit lens model with a singular isothermal ellipsoid results in large errors in the image positions. A significantly improved fit is obtained after the addition of a shear component, suggesting that the lensing system is more complex and may consist of multiple deflectors. The integrated radio spectrum of the background source indicates that it is a GigaHertz-Peaked Spectrum (GPS) source. It may therefore be possible to resolve structure within the radio images with deep VLBI observations and thus better constrain the lensing mass distribution. 
\end{abstract}


\section{Introduction}
Building upon the successful methodology of the Jodrell Bank-VLA Astrometric Survey (JVAS; Patnaik et al. 1992; Browne et al. 1998; Wilkinson et al. 1998) which is known to contain 6 examples of gravitational lensing (King et al. 1999) within a parent sample of 2384 sources, the Cosmic Lens All-Sky Survey (CLASS; for example Browne 1999) aims to find new examples of strong gravitational lensing within a flux-limited sample of $\sim12\,000$ flat-spectrum radio sources observed with the VLA in A configuration at 8.4~GHz. Upon completion of CLASS the lensing frequency within the combined samples of JVAS and CLASS will allow constraints to be placed on the cosmological constant $\Omega_{\Lambda}$. This has already been achieved for the JVAS sample (Helbig et al. 1999). Also, individual lens systems can be used to determine the Hubble parameter (Refsdal 1964) by measuring the time delay between individual images in a system.

So far 18 lens systems have been found in JVAS and CLASS\footnote{JVAS and CLASS lens survey papers can be found at URL {\tt http://gladia.astro.rug.nl:8000/ceres/papers/papers.html}.}. Prior to the final group of CLASS observations which are currently being followed up with MERLIN and the VLBA, this paper reports on the discovery of a new lens system, B0128+437, originally observed in the first phase of CLASS. The system was found after a re-analysis of the entire CLASS dataset.

\section{VLA Re-analysis and MERLIN Observations}
In order to ensure uniformity over all epochs of observations, the entire CLASS dataset was re-flagged and re-calibrated together in a standard way within AIPS. The data were then re-mapped using an automatic script within DIFMAP (Shepherd 1997; Pearson et al. 1994) and Gaussian model components were fitted to the components in each map. All sources with multiple compact model components with separations in the range 0\farcs3 to 6\arcsec and flux density ratios $\le$10:1 were selected. The beam size of the VLA at 8.4~GHz in A configuration is $\sim$0\farcs2, and a compact component is defined to have a Gaussian diameter (FWHM) $\le$170~mas. Also, only those systems where the total flux density in all components added together is $\ge$20~mJy were selected. The vast majority of lens candidates found this way have been followed up during previous phases of CLASS. The small number of {\em new} candidates arising from the re-analysis and re-selection process were followed up with MERLIN at 5~GHz.

\begin{figure}
\begin{center}
\setlength{\unitlength}{1cm}
\begin{picture}(6,8)
\put(-4.2,10.0){\includegraphics{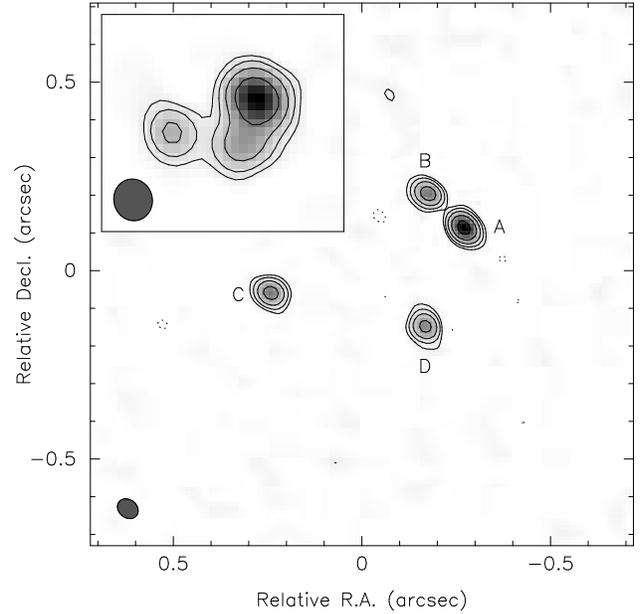}}
\end{picture}
\caption{MERLIN 5~GHz map of B0128+437 taken on 2000 March 09 centered on (J2000.0) R.A. 01$^{\mbox{\scriptsize h}}$31$^{\mbox{\scriptsize m}}$13\fs420 dec. +43\degr58\arcmin13\farcs02. The contours are at -3, 3, 6, 12, 24, and 48 times the rms noise of 277~$\mu$Jy~beam$^{-1}$. The beam size is 58.8~mas $\times$ 47.7~mas in position angle 50\fdg9. The top-left panel shows the VLA 8.4~GHz CLASS map taken on 1994 March 05 for comparison. Contours are at -3, 3, 6, 12, and 24 times the rms noise of 271~$\mu$Jy~beam$^{-1}$. The beam size is 0\farcs246 $\times$ 0\farcs225 in position angle 14\fdg7. Note that the scale of the top-left panel is $\sim0.5$ that of the main map.}\label{merlinmap}
\end{center}
\end{figure}

The top-left panel of Figure \ref{merlinmap} shows the VLA 8.4~GHz map of B0128+437 after the re-calibration of the CLASS data, obtained on 1994 March 05 as part of CLASS ``phase 1''. The maximum component separation is 534~mas which is less than three times the beam size and as a result only three components are resolved. It can be seen {\it a posteriori} that this map is consistent with a four image lens system where the brightest component in the VLA map consists of an unresolved blend of two images. The total flux density of the system is 24.6~mJy and the flux density ratio between the brightest and faintest components is $\sim$3.75:1. The details of the three components are given in Table \ref{vladetails}.

\begin{table}
\begin{center}
\caption{Image parameters for B0128+437 derived from the Gaussian components fitted to the VLA 8.4~GHz data. Image positions are offset from component A at (J2000.0) R.A. 01$^{\mbox{\scriptsize h}}$31$^{\mbox{\scriptsize m}}$13\fs471 dec. +43\degr58\arcmin12\farcs938.}\label{vladetails}
\begin{tabular}{lccrr}
 & \multicolumn{2}{c}{Offset (arcsec)}& & \\
\vspace{0.1cm}
Component & East & North & $S_{8.4}$ (mJy)\\
A+B & +0.000 & +0.000 & 14.8 \\
C & +0.497 & $-$0.188 & 3.9  \\
D & +0.076 & $-$0.266 & 5.8  \\
\end{tabular}
\end{center}
\end{table}

B0128+437 was observed at 5~GHz on 2000 March 09 by MERLIN as part of `snapshot' observations of the new lens candidates from the CLASS VLA data re-analysis. The data were edited and calibrated using the standard MERLIN programs and AIPS, and then mapped to an rms noise level of 277~$\mu$Jy beam$^{-1}$ with a beam size of $\sim$50~mas using DIFMAP. Four components were detected (Figure \ref{merlinmap}) with a maximum component separation of 542~mas. The brightest component in the VLA map is well resolved into two components in the MERLIN map with a separation of 136~mas. The positions of all four components are consistent with those expected from a four-image lens system. The individual components have been modelled with four Gaussian components giving a total flux density of 47.7~mJy. All four components are unresolved i.e. $<30$~mas to within the errors and modelling the source with four $\delta$ functions or four Gaussian components resulted in similar values of $\chi^2$. Details of the model components are given in Table \ref{merlindetails}. Components are labelled (anti-clockwise from the brightest) A to D.

The flux density ratios between components appear to have slightly different values at 8.4~GHz and 5~GHz. However, components A, B, and D are contained within an area only two beam-widths across in the VLA 8.4~GHz map and significant errors in the flux density of each component can therefore arise when the map is deconvolved with the beam. Furthermore, the $3\sigma$ level at each of the two frequencies is $\sim$1~mJy~beam$^{-1}$. The combination of these two factors means that the differences between the flux density ratios at each frequency are not significant.

\begin{table}
\begin{center}
\caption{Image parameters for B0128+437 derived from the Gaussian components fitted to the MERLIN 5~GHz data. Image positions are offset from component A at (J2000.0) R.A. 01$^{\mbox{\scriptsize h}}$31$^{\mbox{\scriptsize m}}$13\fs405 dec. +43\degr58\arcmin13\farcs14.}\label{merlindetails}
\begin{tabular}{lccrr}
 & \multicolumn{2}{c}{Offset (arcsec)}& & \\
\vspace{0.1cm}
Component & East & North & $S_{5}$ (mJy) \\
A & +0.000 & +0.000 & 18.9 \\
B & +0.098 & +0.094 & 9.5 \\
C & +0.520 & $-$0.172 & 10.1 \\
D & +0.108 & $-$0.250 & 9.2 \\
\end{tabular}
\end{center}
\end{table}

\section{Lens Galaxy Modelling}
\label{modelling}
\begin{figure}
\begin{center}
\setlength{\unitlength}{1cm}
\begin{picture}(6,8)
\put(-2,-0.65){\includegraphics{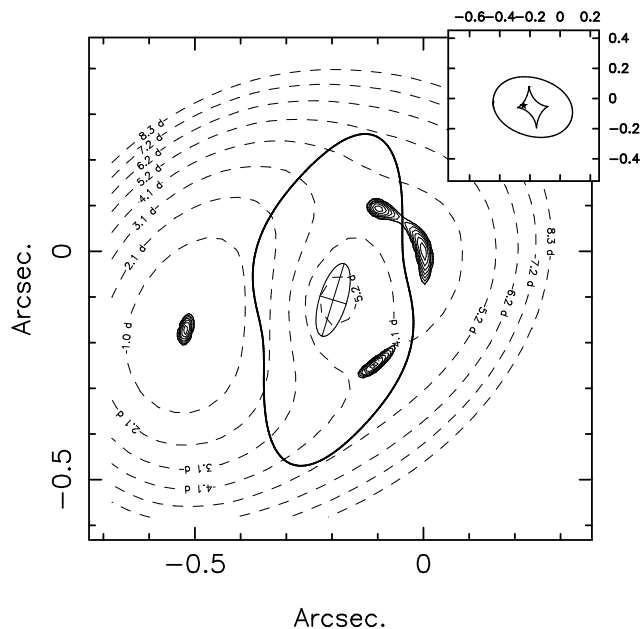}}
\end{picture}
\caption{An attempt at modelling B0128+437 with external shear. The model images are shown as solid contours, whilst the time-delay surface is shown as dashed contours. The solid line shows the critical curve while the top-right sub-image shows the caustics where multiple images are formed. The source position is shown in the sub-image as a star to the left of the centre of the lensing galaxy, just inside the inner caustic of the system. The central cross-haired ellipse indicates the galaxy position and orientation.}\label{modelmap}
\end{center}
\end{figure}

An attempt has been made to model the B0128+437 lens system with a singular isothermal ellipsoid mass distribution (for example Kormann, Schneider \& Bartelmann 1994). The four MERLIN image flux densities and positions have been used as input parameters, providing eleven constraints (eight from image positions and three from image flux ratios). The model parameters which were allowed to vary were the source and lens positions, the lens velocity dispersion, ellipticity and position angle. Thus there were seven free parameters, and hence four degrees of freedom. A 3~mas error in the relative positions of each component and a 20~per~cent error in the flux densities to take account of source variability and systematic errors were allowed. A lens redshift of $z_l=0.5$, source redshift of $z_s=1.5$, and a flat universe with $\Omega_{M}=1.0$ were assumed. The resulting model suggests that the lens mass has an axial ratio $b/a\sim0.76$. However, the differences in the positions of the four model images from those in the actual map are all in the range 30 to 100~mas which clearly shows that a singular isothermal ellipsoid model does not reproduce the observations well. This suggests that the system may consist of a complex deflector and may involve other nearby galaxies.

Modelling with the addition of an external shear component resulted in image positions and flux ratios being reproduced well (see Figure \ref{modelmap}) with image positions within 1~mas and flux ratios within a few per~cent of the observed values; the shear has a value of $\sim$0.23. This result also suggests that there is an additional lensing component external to the main lensing galaxy to the north-east or south-west. Optical and/or infrared imaging must now be carried out to determine the positions and numbers of possible deflectors before more accurate modelling can be achieved. No optical counterpart has been found at the position of the radio images on POSS-II, placing a limit of $m_R>20$ on the main lensing galaxy.

The image configuration in B0128+437 is similar to that of another CLASS lens system, B1555+375 (Marlow et al. 1999). This system has been mapped with greater sensitivity than B0128+437 and has a well-constrained mass model for the lensing galaxy. From the model for B1555+375, the magnification of individual images ranges from 0.5 to 8.2, and the combined magnification for all four images is $\sim$20. This comparison is used in the next section.

\section{The Nature of the Radio Source in B0128+437}
\begin{figure}
\begin{center}
\setlength{\unitlength}{1cm}
\begin{picture}(6,6.0)
\put(8,-1){\includegraphics{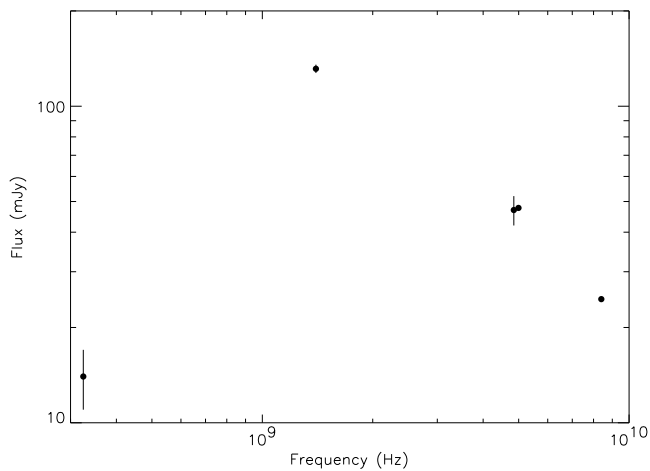}}
\end{picture}
\caption{The radio spectrum of B0128+437. The data have been obtained from the following surveys (with epoch) from 325~MHz to 8.4~GHz: WENSS (05/93); NVSS (12/93); GB6 ($\sim$/87); CLASS (03/00); CLASS (03/94). Error bars are shown at 1$\sigma$.}\label{spectrum}
\end{center}
\end{figure}

The integrated spectrum of the background source in B0128+437 can be obtained from the current MERLIN and VLA observations at 5~GHz and 8.4~GHz, and the GB6 (Gregory et al. 1996), NVSS (Condon et al. 1998), and WENSS (Regelink et al. 1997) surveys at 4.85~GHz, 1.4~GHz and 325~MHz respectively. A plot of the radio spectrum from these data is shown in Figure \ref{spectrum}. Despite the sparsity of points, the spectrum clearly peaks at $\sim$1~GHz which shows that the source is a GigaHertz Peaked Spectrum (GPS) source. 

Snellen et al. (2000) found that both the turnover frequency $\nu_{\mbox{\scriptsize peak}}$ and the turnover flux density $S_{\mbox{\scriptsize peak}}$ of GPS sources correlates with their overall maximum angular size (component separation, not the maximum size of a {\em single} GPS source component) over a flux density range for the turnover ranging from $\sim$40~mJy to $\sim$5~Jy. It is estimated that $\nu_{\mbox{\scriptsize peak}}$ for B0128+437 is $\sim$1~GHz with a flux density $S_{\mbox{\scriptsize peak}}\sim$100~mJy. However, the source is being magnified and lensed into four images. Assuming that the total magnification in B0128+437 is similar to that in B1555+375 (see Section \ref{modelling}), the `real' turnover flux density of B0128+437 is estimated to be $\sim$5~mJy. The observed near-linear correlation between the overall maximum angular size and $S_{\mbox{\scriptsize peak}}^{1/2}\nu_{\mbox{\scriptsize peak}}^{-5/4}$ suggests that the intrinsic angular size of the radio source in B0128+437 is between 1 and 10~mas. Since the source is being magnified by up to $\sim$10 times in the strongest image the angular size of the source could increase to between 3 and 30~mas in this instance. The fact that the strongest image is unresolved in the MERLIN 5~GHz map also places an upper limit of $\sim$30~mas on the apparent angular size of the source in this image. VLBI observations at 1.4~GHz with a resolution of a few milliarcseconds might be able to resolve the images, which would place additional constraints on the lens model.

The prospect of using this system to measure the Hubble parameter is unfortunately not promising. If the radio source is indeed a GPS source, it will be a member of one of the least variable classes of extragalactic radio source (Aller, Aller \& Hughes 1992). The GB6 survey data show that the 4.85~GHz flux density was $47\pm5$~mJy between 1986 November and 1987 October. The MERLIN observations at 5~GHz on 2000 March give a total flux density of $47.7\pm0.3$~mJy. While this does not rule out the possibility of detecting variability within the source, it is consistent with the source being non-variable.

\section{Small Quad Lenses}
Since, in the VLA 8.4~GHz CLASS map of B0128+437 images A and D were not resolved, as was also the case with B1555+375, other compact four-image lens systems may have been missed in the search of JVAS and CLASS if two of the images had a separation smaller than the beam size. In such cases a single extended Gaussian component would have been fitted to these images during the automatic mapping process. The lens candidate selection criterion stipulating multiple {\em compact} components would then result in rejection of such a system as a lens candidate unless the remaining two images were detected and fitted with compact Gaussian components. In order to check that no such systems have been missed, a new set of selection criteria based on the expected image configurations of small-quad lens systems was devised. Five new candidates were found, all of which have been followed-up with MERLIN at 5~GHz. In each case, the `resolved' component in the VLA 8.4~GHz CLASS map was shown by MERLIN to be a single extended component. All five candidates were rejected as lenses based on surface brightness and morphological arguments. It is unlikely that there are additional lens systems like B0128+437 or B1555+375 within the JVAS and CLASS samples.

\section{Summary and Future Work}
A new gravitational lens system, B0128+437, has been discovered in the course of a re-analysis of the entire CLASS dataset. The system consists of four images of a background radio source which appears to be a GPS source. It might be possible to detect extended structure if the system is observed with VLBI. Initial attempts to model the lens mass with a singular isothermal ellipsoid distribution have proved unsuccessful, indicating that there may be multiple deflectors contributing to the lens.

The discovery of B0128+437 has increased the number of known compact lens systems. Whilst there is no optical information to constrain the lensing mass, Turner, Ostriker \& Gott (1984) suggest that lower separation lenses are more likely to be due to spiral galaxies. At present the mass distributions of these systems are not well understood at cosmological lensing redshifts (for example Keeton \& Kochanek, 1998) making B0128+437 of particular interest.

In order to constrain the lens mass model, further radio observations with MERLIN and VLBI must be made to better determine the image flux densities and positions. The detection of extended structure within the source will provide further parameters with which to constrain the lens mass model. Optical and/or infrared imaging to locate the lensing mass(es), and spectroscopy to find the lens and source redshifts are also needed for a satisfactory lens model.

\section*{Acknowledgments}
The VLA is the Very Large Array and is operated by the National Radio Astronomy Observatory which is a facility of the National Science Foundation operated under cooperative agreement by Associated Universities, Inc. MERLIN is the Multi-Element Radio Linked Interferometer Network and is a national facility operated by the University of Manchester on behalf of PPARC.  This research was supported in part by the European Commission, TMR Programme, Research Network Contract ERBFMRXCT96-0034 `CERES'. PMP would also like to thank PPARC for the support of a studentship award.

\label{lastpage}

\end{document}